\documentclass[letter]{aa} 
\usepackage{amsmath}
\usepackage{graphicx}
\usepackage{graphics}
\usepackage{subfigure}
\usepackage{txfonts}
\usepackage{natbib}
\usepackage{xcolor}

\bibpunct{(}{)}{;}{a}{}{,}

\def\teff{\ifmmode T_{\rm eff} \else $T_{\mathrm{eff}}$\fi}

\def\ltsima{$\buildrel<\over\sim$}
\def\lsim{\lower.5ex\hbox{\ltsima}}

\newcommand{\hii}{H~{\sc ii}}
\newcommand{\ha}{\ifmmode {\rm H}\alpha \else H$\alpha$\fi}
\newcommand{\hb}{\ifmmode {\rm H}\beta \else H$\beta$\fi}
\newcommand{\lya}{\ifmmode {\rm Ly}\alpha \else Ly$\alpha$\fi}

\newcommand{\ebv}{\ifmmode E_{\rm B-V} \else $E_{\rm B-V}$\fi}
\newcommand{\av}{\ifmmode A_{\rm V} \else $A_{\rm V}$\fi}
\def\micron{$\mu$m}

\def\msun{\ifmmode M_{\odot} \else M$_{\odot}$\fi}
\def\msunyr{\ifmmode M_{\odot} {\rm yr}^{-1} \else M$_{\odot}$ yr$^{-1}$\fi}
\def\zsun{\ifmmode Z_{\odot} \else Z$_{\odot}$\fi}

\def\lsun{\ifmmode L_{\odot} \else L$_{\odot}$\fi}

\def\mup{\ifmmode M_{\rm up} \else M$_{\rm up}$\fi}
\def\mlow{\ifmmode M_{\rm low} \else M$_{\rm low}$\fi}

\newcommand{\oh}{\ifmmode 12 + \log({\rm O/H}) \else$12 + \log({\rm
O/H})$\fi}
\newcommand{\nii}{[N~{\sc ii}]}

\newcommand{\cii}{[C~{\sc ii}]}

\def\flyf{\ifmmode f_{\rm Lyf} \else $f_{\rm Lyf}$\fi}
\def\pz{\ifmmode P(z) \else $P(z)$\fi}
\def\ki2{\ifmmode \chi^2 \else $\chi^2$\fi}
\def\zphot{\ifmmode z_{\rm phot} \else $z_{\rm phot}$\fi}

\newcommand{\xphot}{\ifmmode x_\gamma \else $v_\gamma$\fi}
\newcommand{\xobs}{\ifmmode x_{\rm obs} \else $x_{\rm obs}$\fi}
\newcommand{\xcmf}{\ifmmode x_{\rm CMF} \else $x_{\rm CMF}$\fi}
\newcommand{\vexp}{\ifmmode V_{\rm exp} \else $V_{\rm exp}$\fi}
\newcommand{\vmax}{\ifmmode V_{\rm max} \else $V_{\rm max}$\fi}
\newcommand{\nh}{\ifmmode N_{\rm HI} \else $N_{\rm HI}$\fi}
\newcommand{\dv}{\ifmmode \Delta v({\rm em-abs}) \else $\Delta v({\rm em}-{\rm abs})$\fi}

\def\fesc{\ifmmode f_{\rm esc} \else $f_{\rm esc}$\fi}

\def\frellya{\ifmmode f^{\rm rel}_{\rm{Ly}\alpha} \else $f^{\rm rel}_{\rm{Ly}\alpha}$\fi}

\newcommand{\mstar}{\ifmmode M_\star \else $M_\star$\fi}
\newcommand{\mdust}{\ifmmode M_d \else $M_d$\fi}
\newcommand{\muv}{\ifmmode M_{1500} \else $M_{1500}$\fi}
\newcommand{\luv}{\ifmmode L_{\rm UV} \else $L_{\rm UV}$\fi}
\newcommand{\lir}{\ifmmode L_{\rm IR} \else $L_{\rm IR}$\fi}
\newcommand{\lfir}{\ifmmode L_{\rm FIR} \else $L_{\rm FIR}$\fi}
\newcommand{\lbol}{\ifmmode L_{\rm bol} \else $L_{\rm bol}$\fi}
\newcommand{\liruv}{\ifmmode L_{\rm IR+UV} \else $L_{\rm IR+UV}$\fi}
\newcommand{\liroveruv}{\ifmmode L_{\rm IR}/L_{\rm UV} \else $L_{\rm IR}/L_{\rm UV}$\fi}
\newcommand{\nlyc}{\ifmmode N_{\rm Lyc} \else $N_{\rm Lyc} $\fi}
\newcommand{\rholyc}{\ifmmode \rho_{\rm Lyc} \else $\rho_{\rm Lyc} $\fi}
\newcommand{\auv}{\ifmmode  A_{\rm UV} \else $A_{\rm UV}$\fi}

\newcommand{\Cii}{[C~{\sc ii}] 158\micron}
\newcommand{\lcii}{\ifmmode L_{[\rm CII]} \else $L_{[\rm CII]}$\fi}
\newcommand{\lco}{\ifmmode L_{\rm CO} \else $L_{\rm CO}$\fi}
\newcommand{\lirngc}{\ifmmode L_{\rm IR}^{\rm N6946} \else $L_{\rm IR}^{\rm N6946}$\fi}

\newcommand{\macs}{MACS J0451+0006}

\begin{document}
    \title{ALMA detection of \Cii\ emission from a strongly lensed $z=2.013$ star-forming galaxy\thanks{Based on ALMA observations 2011.0.00130.S.}}  
\subtitle{}
  \author{D. Schaerer\inst{1,2},   F. Boone\inst{2},
  T. Jones\inst{3},
M. Dessauges-Zavadsky\inst{1},
  P. Sklias\inst{1}, M. Zamojski\inst{1}, A. Cava\inst{1},
  J. Richard\inst{4},
  R. Ellis\inst{5},
  T. D. Rawle\inst{6},
  E. Egami\inst{7},
  F. Combes\inst{8}}
  \institute{
Observatoire de Gen\`eve, D\'epartement d'Astronomie, Universit\'e de Gen\`eve, 51 Ch. des Maillettes, 1290 Versoix, Switzerland
         \and
CNRS, IRAP, 14 Avenue E. Belin, 31400 Toulouse, France
	\and	
Department of Physics, University of California, Santa Barbara, CA 93106, USA
	\and	
CRAL, Observatoire de Lyon, Universit\'e de Lyon 1, 9 avenue Ch. Andr\'e, F-69561 Saint-Genis Laval, France
	 \and
Department of Astronomy, California Institute of Technology, MS 249-17, Pasadena, CA 91125, USA
	 \and
ESAC, ESA, PO Box 78, Villanueva de la Ca\~nada, Madrid 28691, Spain
	\and
Steward Observatory, University of Arizona, 933 N. Cherry Ave, Tucson, AZ 85721, USA
	\and
Observatoire de Paris, LERMA, 61 Av. de l'Observatoire, 75014, Paris, France 
}

\authorrunning{D. Schaerer et al.}
\titlerunning{ALMA detection of \Cii\ emission from a strongly lensed $z=2$ star-forming galaxy}

\date{Accepted for publication in Astronomy \& Astrophysics Letters}

\abstract{}
{Our objectives are to determine the properties of the interstellar medium (ISM) and of star-formation in typical star-forming galaxies
at high redshift.}
{Following up on our previous multi-wavelength observations with HST, Spitzer, Herschel, and the Plateau de Bure Interferometer (PdBI),
we have studied a strongly lensed $z=2.013$ galaxy, the arc behind the galaxy cluster \macs, with ALMA to measure the 
\Cii\ emission line, one of the main coolants of the ISM.}
{\cii\ emission from the southern part of this galaxy is detected at $10 \sigma$. Taking into account strong gravitational lensing,
which provides a magnification of $\mu=49$, the intrinsic lensing-corrected \cii\ luminosity is $\lcii=1.2 \times 10^8$ \lsun. 
The observed ratio of \cii-to-IR emission, $\lcii/\lfir \approx(1.2-2.4) \times 10^{-3}$, is found to be similar to that
in nearby galaxies. The same also holds for the observed ratio \lcii/\lco $=2.3 \times 10^3$, which is comparable to that of star-forming 
galaxies and active galaxy nuclei (AGN) at low redshift.
}
{We utilize strong gravitational lensing to extend diagnostic studies of 
the cold ISM to an order of magnitude lower luminosity ($\lir \sim (1.1-1.3) \times 10^{11}$ \lsun)
and SFR than previous work at high redshift. While larger samples are needed, our 
results provide evidence that the cold ISM of typical high redshift 
galaxies has physical characteristics similar to normal star forming 
galaxies in the local Universe.}

 \keywords{Galaxies: high-redshift -- Galaxies: starburst -- Galaxies: ISM --  Galaxies: formation --  Radio lines: galaxies}

  \maketitle

\section{Introduction}
\label{s_intro}
Observations of CO molecular emission and fine structure lines, such as \Cii, provide interesting insight into the gas reservoir
and interstellar medium (ISM) of galaxies \citep{Carilli2013Cool-Gas-in-Hig}.
Although such observations are now yielding first measures of the molecular gas fraction at high redshifts 
\citep[e.g.][]{2010Natur.463..781T,Saintonge2013Validation-of-t}
and hints on the ISM (photo-dissociation regions (PDRs), and \hii\ regions) up to $z \sim 6$ 
\citep{Riechers2013A-dust-obscured,De-Breuck2014ALMA-resolves-t,Rawle2014C-II-and-12CO1-,Riechers2014ALMA-Imaging-of},
very little is known about the properties of the most abundant galaxies with relatively low IR luminosities
and modest stellar masses. Gravitationally lensed sources offer a unique chance to sample this regime.

The Herschel Lensing Survey of massive galaxy clusters \citep{Egami2010The-Herschel-Le} and accompanying 
multi-wavelength observations were in particular designed
to constrain the properties of  typical-luminosity galaxies
by probing beyond nominal sensitivity limits with the power of strong gravitational lensing.
Detailed stellar, star-formation, and dust properties of a sample of seven galaxies at $z \sim 1.5-3$ have been determined 
by \cite{Sklias2014Star-formation-}.
Using CO observations we have also studied their molecular gas properties in 
\cite{Dessauges-Zavadsky2014Molecular-gas-c}.
In cycle 0 we have been able to observe one of these galaxies with ALMA, the gravitationally lensed, multiply-imaged arc \macs\ at $z=2.013$,
to study its ISM/PDR properties,
as part of an ongoing ALMA program to study line emission in high redshift gravitationally-lensed galaxies (PI: Ellis).
We here present results concerning the \Cii\ emission, providing some of the first time information on \cii\ and CO in an IR-faint 
($\lir \sim (1.1-1.4) \times 10^{11}$ \lsun), low mass ($\mstar \sim 2.5 \times 10^9$ \msun\ for a Chabrier IMF) star-forming galaxy at high redshift.

The observational data are described in Sect.\ \ref{s_obs}. Our main results are presented and discussed in Sect.\ \ref{s_results}.
Section \ref{s_conclude} summarizes our main conclusions.
We adopt a $\Lambda$-CDM cosmological model with $H_{0}$=70 km s$^{-1}$ Mpc$^{-1}$, 
$\Omega_{m}$=0.3 and $\Omega_{\Lambda}$=0.7.

\section{Observations}
\label{s_obs}

\begin{figure}[htb]
\centering
\includegraphics[width=8.8cm]{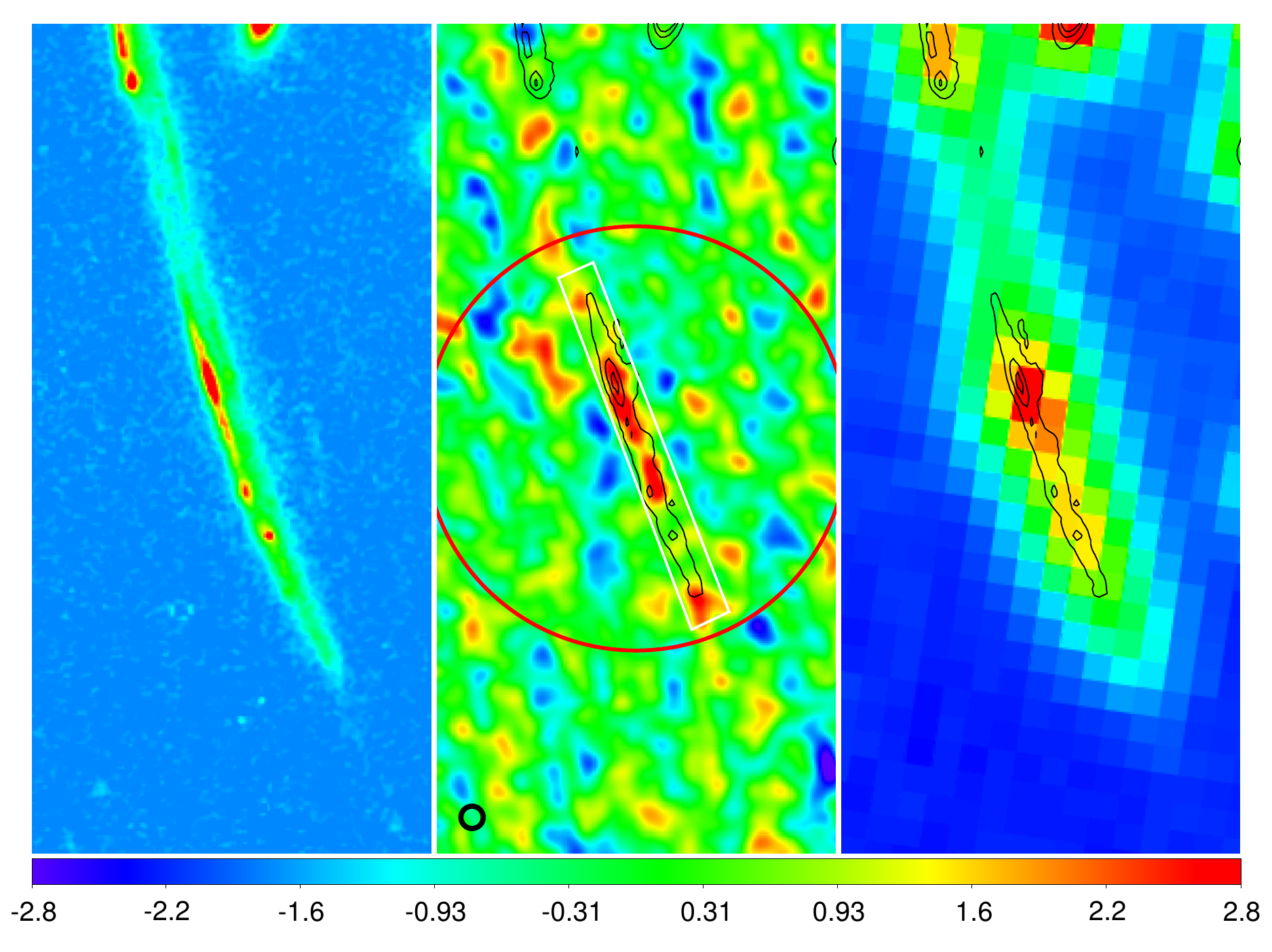}
\caption{{\em From left to right}: HST/WFC3 image in the filter F140W, ALMA \Cii\  integrated map
not corrected for primary beam attenuation, and {\it Spitzer} map with IRAC at 3.6\,$\mu$m. 
Contours in the middle and right panel show the HST flux. The color scale of the ALMA map is shown at the bottom in Jy\,km\,s$^{-1}$.
To emphasize the emission from the arc, the ALMA data were tapered and the resolution is 0.5$''$ (represented by the black circle at the bottom left). The red circle represents the ALMA primary beam 
at half its maximum, its diameter is 9.5\arcsec; the white box delineates the region over which the 
continuum and the spectra have been integrated. The images are 9\arcsec\ by 18\arcsec\ side with a standard orientation (N up, E left).
}
\label{fig_map}
\end{figure}

\subsection{ALMA observations and results}
The \macs\ arc was targeted with ALMA in cycle 0 to map \cii\
emission on $\sim 200$ pc scales. Our aims are to measure the spatial
distribution and kinematics of the cold ISM/PDRs, and to compare this with
the distribution and kinematics of star forming regions traced by
H$\alpha$ (Jones et al. 2010). 
Although the delivered data did not meet our requested sensitivity,
it is sufficient to examine the average ISM/PDR properties from the integrated \cii\ emission.

Our source was observed by 25 antennas in band 9 during a $\sim$1h track including calibration 
scans on the 31st of December 2012.
Projected baseline lengths range from 13 to 376 meters.
The correlator was set up to obtain 4 spectral sub-bands of 2\,GHz with 128  channels of 15.625\,MHz. 
The receivers were tuned to center the bands on the following frequencies in GHz: 628.826, 630.520, 632.220  and 648.234.
Passband, flux and phase calibrations used the quasar J0423-013, extrapolating from ALMA band 3 ($\sim$100\,GHz) and band 7 ($\sim$300\,GHz) observations made between December 14-16, 2012, which leads to a relatively large uncertainty for the flux calibration, of the order of $\sim$ 25\%. 
The visibilities were calibrated and the image cleaned with the CASA software. With natural weighting the clean beam full width at half maximum (FWHM) is $0.33''\times 0.34''$ and the noise RMS is $\sigma=9.8$\,mJy\,beam$^{-1}$ in 14.9\,km\,s$^{-1}$ channels.

Flagging bad channels and masking out the \Cii\ line we integrate the 4 sub-bands to produce a continuum map, 
obtaining a noise of $\sigma_{\rm cont}=0.84$\,mJy\,beam$^{-1}$.
No continuum emission is detected at the position of the gravitational arc.
Integrating the flux in a box encompassing the arc as shown in the Fig.\,\ref{fig_map} and correcting for the primary beam, we obtain 
a 3 $\sigma$ upper limit of $<27$ mJy, 
which is compatible with the SPIRE 500 \micron\ band (Zamojski et al., in preparation).

\begin{figure}[htb]
\centering
\includegraphics[width=8.8cm]{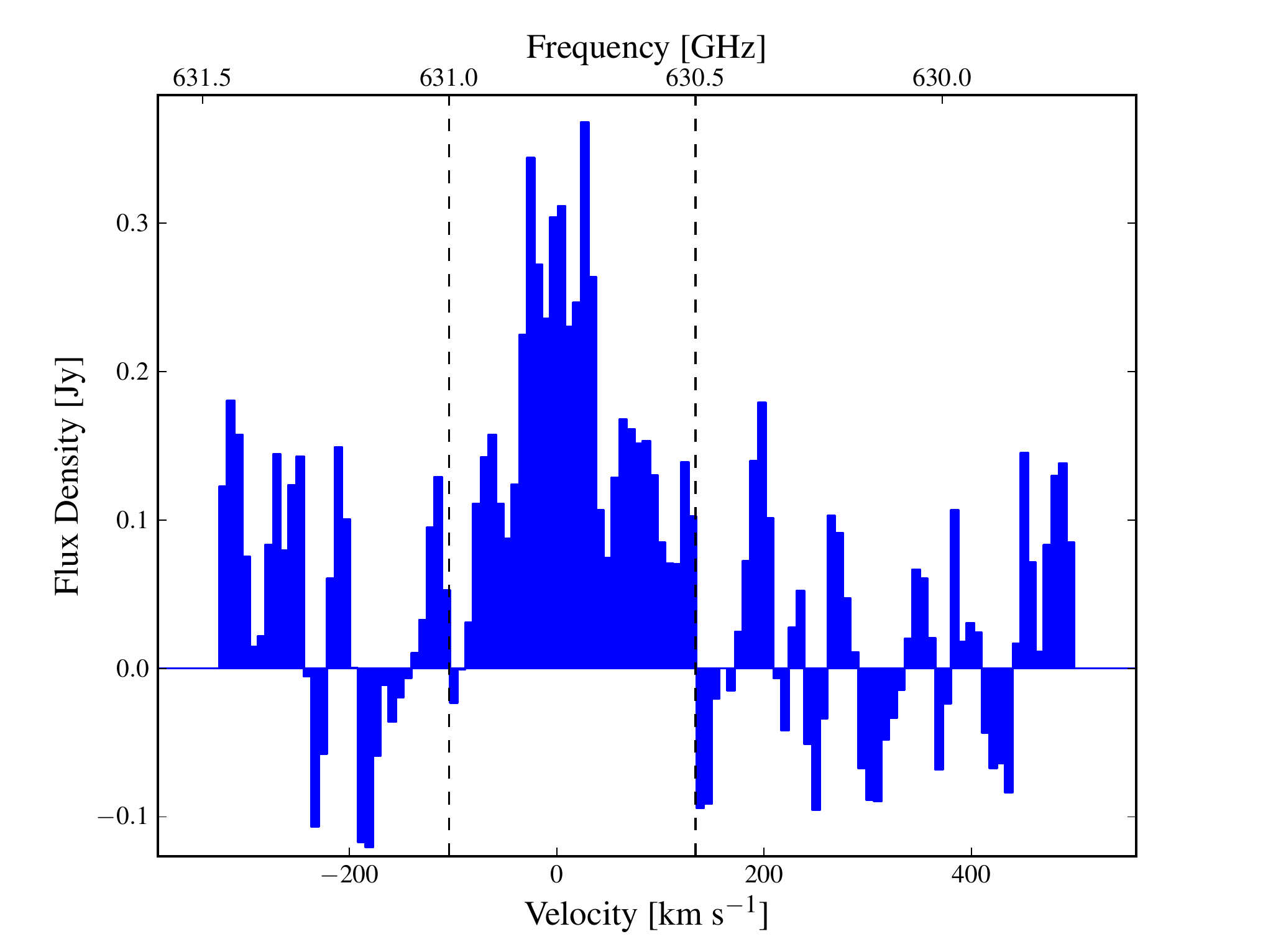}
\caption{Observed \Cii \ spectrum obtained by integrating the datacube corrected for the primary beam in a box encompassing the southern part of the arc. The dashed vertical lines show the window within which the channels are summed to obtain the line flux. The Doppler velocities (lower axis) are computed with respect to the \Cii\ frequency at $z=2.013$, namely 630.78\,GHz.}
\label{fig_cii_line}
\end{figure}

Integrating the spectra corrected for the primary beam in the same spatial box we clearly detect the \Cii\ line shown in Fig.\ \ref{fig_cii_line}.
The flux integrated in the range 630.5-631.0\,GHz is $S_{\rm CII}=37.7 \pm 3.7$ Jy\,km\,s$^{-1}$,
which corresponds to an intrinsic luminosity of $\lcii = 1.2 \times 10^8$ \lsun, 
after correction for magnification by a factor $\mu=49$. 
This magnification corresponds both to the mean magnification of the total arc \citep[cf.][]{Jones2010Resolved-spectr} 
and to the mean over the region concerned here, as determined from an updated version of the lensing model by \cite{Richard2011The-emission-li} 
and recent updates to it. We estimate an uncertainty of $\mu=49 \pm 5$.
Spatially, the \cii\ emission follows closely the arc, as traced by the 1.4 and 3.6 \micron\ emission. 
Given that the \cii\ emission is very compact along the direction perpendicular to the arc, the flux is thus
well constrained by the baselines in that direction. 
Furthermore, the length of the arc is smaller than the primary beam and 
the corresponding baselines are well sampled.
Therefore, the flux lost due to missing short spacings should 
be small, and negligible compared to the flux uncertainties quoted above.

\begin{table}[htb]
\caption{Measured \Cii\ and other key properties of the strongly lensed arc in \macs. 
The intrinsic luminosities are determined adopting the magnification factor $\mu=49$ from the updated 
lens model of \cite{Richard2011The-emission-li}.}
\begin{tabular}{llllllllllllll}
\hline
\hline
Quantity & value \\
\hline
$z^a$ & 2.013 \\
$\nu_{\rm CII}$ & 630.743 $\pm$ 0.015 GHz\\
$S_{\rm CII}^b$ & $37.7 \pm 3.8$ Jy km/s \\
$\lcii \times \mu$ & $ (5.94\pm 0.6) \times 10^{9}$ \lsun \\
$\lcii^b$ & $(1.21 \pm 0.12) \times 10^{8}$ \lsun \\
\lir$^c$ &  $(1.1-1.3) \times 10^{11}$ \lsun \\
\lfir$^d$ &  $(0.5-0.9) \times 10^{11}$ \lsun \\
\lco$^e$ &  $(5.1 \pm 1.5) \times 10^4$ \lsun\\
\hline  
\label{t_derived}
\end{tabular}
\vspace{-0.5cm}
\tablefoot{$^a$ \ha\ redshift from \cite{Richard2011The-emission-li}.
$^b$ Formal error, excluding  $\sim$ 25\% systematic uncertainty from flux calibration.
$^c$ IR luminosity of the S of the arc \cite[cf.][]{Sklias2014Star-formation-}.
$^d$ Range of \lfir\ characterizing uncertainty in spatial separation of N and S arc.
$^e$ From \cite{Dessauges-Zavadsky2014Molecular-gas-c}.
}
\end{table}

\subsection{Other observations}
The CO(3-2) line was detected with the PdBI
by \cite{Dessauges-Zavadsky2014Molecular-gas-c}.
The emission originates from the southern part of the arc, i.e.\ the same region as targeted with ALMA.
The CO flux translates to a lensing-corrected CO(1-0) luminosity of $\lco =(5.1 \pm 1.5) \times 10^4$ \lsun\
assuming a factor $r_{3,1}=0.57$ between the CO(3-2) and CO(1-0) line luminosities, and $\mu=49$
\citep{Dessauges-Zavadsky2014Molecular-gas-c}.
The IR spectral energy distribution of the \macs\ arc, based on Spitzer and Herschel observations, has been studied in detail by \cite{Sklias2014Star-formation-}, 
from which we obtain the IR luminosity for the southern part of the arc\footnote{From the difference between the total luminosity and that of the subdominant northern part.} corresponding to the region of the CO and \cii\ 
detection.
The CO and IR luminosities are listed in Table \ref{t_derived}.
As commonly used, \lfir \ is defined as the luminosity between 42.5 and 122.5 \micron\ rest-frame,
whereas the total IR luminosity, denoted \lir\ here, refers to the 8-1000 \micron\ domain.

Although both the CO and \cii\ emission originate from the region, the two lines show different velocity profiles: 
the former a double-peaked profile and the latter a single component centered on zero velocity. 
The apparent difference between the line profiles could result from the ISM properties of the galaxy
or be due to the relatively low S/N of the CO line. Future ALMA observations should settle this question.

\section{Results}
\label{s_results}

\subsection{Carbon emission from a $z=2$, faint LIRG}
We now compare the observed \cii\ luminosity with that of other star-forming galaxies and AGN
detected in the IR, both at low and high redshift.

\begin{figure}[htb]
\centering
\includegraphics[width=8.8cm]{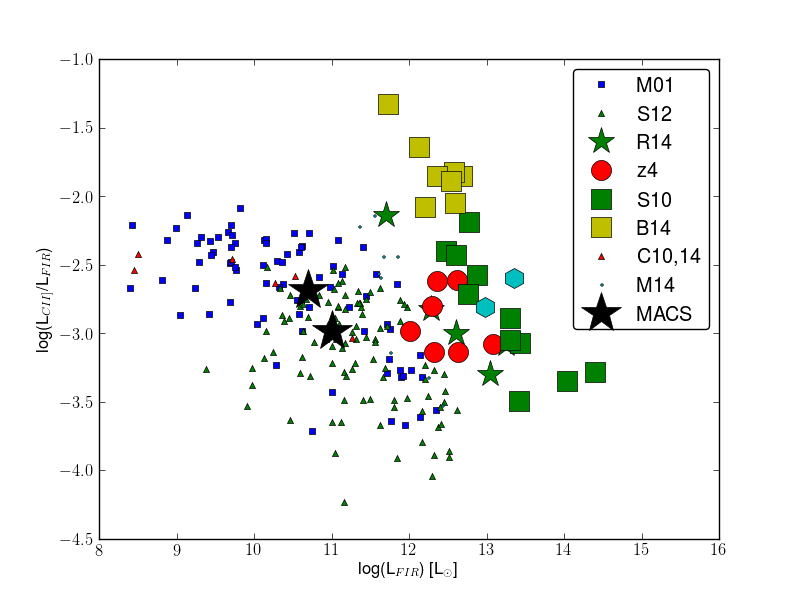}
\caption{\lcii/\lfir\ versus \lfir\ for \macs\ (large, black stars) and comparison samples
which are detected both in \cii\ and in the dust continuum.
For \macs\ the two points are meant to illustrate the uncertainties from \lfir, $\mu$, and the band 9 calibration.
Small symbols show nearby galaxies and AGN, large symbols sources at $z>2$ from the following papers:
M01: \cite{Malhotra2001Far-Infrared-Sp}, 
S12: \cite{2012ApJ...755..171S},
S10: \cite{Stacey2010A-158-mum-C-II-},
B14: \cite{Brisbin2014Strong-C-emissi},
C10,14: \cite{Cormier2010The-effects-of-}, \cite{Cormier2014The-molecular-g},
M14: \cite{Magdis2014A-Far-Infrared-},
R14: \cite{Rawle2014C-II-and-12CO1-},
z4: individual $z>4$ galaxies \citep[cf.][and references therein]{Casey2014Dusty-Star-Form}.
}
\label{fig_cii_lir}
\end{figure}

The \Cii\ luminosity of \macs\ is shown in Fig.\ \ref{fig_cii_lir} as a function of \lfir\
together with previous  detections in nearby and high redshift
galaxies and AGN. Two main results are clearly seen from this figure. 
First, our source has a significantly lower IR luminosity than previously studied galaxies at $z \ga 1-2$.
Indeed, thanks to a high magnification by gravitational lensing our source is currently 
the IR-faintest galaxy detected in \Cii\ at $z>1$, namely a faint LIRG.
Second, relative to \lfir, the \Cii\ emission of \macs\ is $\lcii/\lfir \approx(0.6.-1.2) \times 10^{-3}$, similar to that
in nearby galaxies, as seen e.g.\ by the comparison with the measurements from \cite{Malhotra2001Far-Infrared-Sp}
and \cite{2012ApJ...755..171S}.

Although other \cii\ measurements currently available for high redshift ($z \ga 2$) galaxies are mostly
restricted to the very IR luminous objects ($\lir > 10^{12}$ \lsun)
the relative \lcii/ \lfir\ emission of many of them is also comparable to that of \macs,
as can be seen from Fig.\ \ref{fig_cii_lir}. 
This is also the case for a Lyman break galaxy (LBG) at $z=5.3$ from which \cite{Riechers2014ALMA-Imaging-of} 
have recently detected \cii\ emission with ALMA.
Indeed, although the LBG is undetected in the IR continuum ($\lfir \la 5.3 \times 10^{11}$ \lsun\footnote{$ 3 \sigma$ limit
obtained from their Table 1.}),  \cii\ is fairly strong, corresponding to $\lcii/\lfir > 10^{-2.5}$.
So far, the available data at $z \ga 2$ show a fairly large dispersion in the \lcii/ \lfir\ ratio, 
Some galaxies (marked with B14 in our figure) with high ratios of $\lcii/\lfir \ga 10^{-2}$ were recently found by 
\cite{Brisbin2014Strong-C-emissi}. We note, however, that most of their \cii\ detections are of low significance,
hence associated with large uncertainties.

The large scatter in  \cii/ \lfir\ probably implies that several factors, such as the average radiation field density, 
ISM density, metallicity and others, determine the  \cii/FIR ratio, as amply discussed in the literature 
\citep[e.g.][]{Gracia-Carpio2011Far-infrared-Li,Magdis2014A-Far-Infrared-}.
For the \macs\ arc observed here, we have derived fairly detailed information on its stellar content, star formation rate, 
and dust properties, as well as some measure of its molecular gas 
\citep[cf.][and below]{Sklias2014Star-formation-,Dessauges-Zavadsky2014Molecular-gas-c}.
The dust temperature $T_d$, in particular, has also been determined from our Herschel observations, and
is found to be fairly high, between $\sim$ 50--80 K, for the southern part of the arc observed here.
Although local galaxies show an anti-correlation between the \lcii/\lfir\ ratio
and $T_d$ \citep[cf.][]{Malhotra2001Far-Infrared-Sp,Gracia-Carpio2011Far-infrared-Li,Magdis2014A-Far-Infrared-},
our galaxy shows a ``normal" ratio despite its high dust temperature. A similar result is also found for the
$z=2.957$ lensed galaxy HLSW-01 (SWIRE6) reported by \cite{Magdis2014A-Far-Infrared-}, which has a high
$T_d=51 \pm 2$ K and a normal ratio $\lcii/\lfir=1.5 \times 10^{-3}$.
Clearly, more measurements and additional information on the physical properties of a representative sample
of high-$z$ galaxies are needed to improve our understanding of these objects and their ISM.

Given the agreement of the observed $\lcii/\lfir$ with that of  nearby objects, our galaxy is logically also 
in good agreement with the SFR--\lcii\ relation \citep[cf.][]{2011MNRAS.416.2712D,De-Looze2014The-applicabili}
determined from local star-forming galaxies.
Indeed, from \lir\ the star formation rate is SFR(IR)$=11-13$ \msunyr\  for a Chabrier/Kroupa IMF,
one would obtain $\lcii=(1.5-1.8) \times 10^8$ \lsun, in fair agreement with our observed value.

\begin{figure}[htb]
\centering
\includegraphics[width=8.8cm]{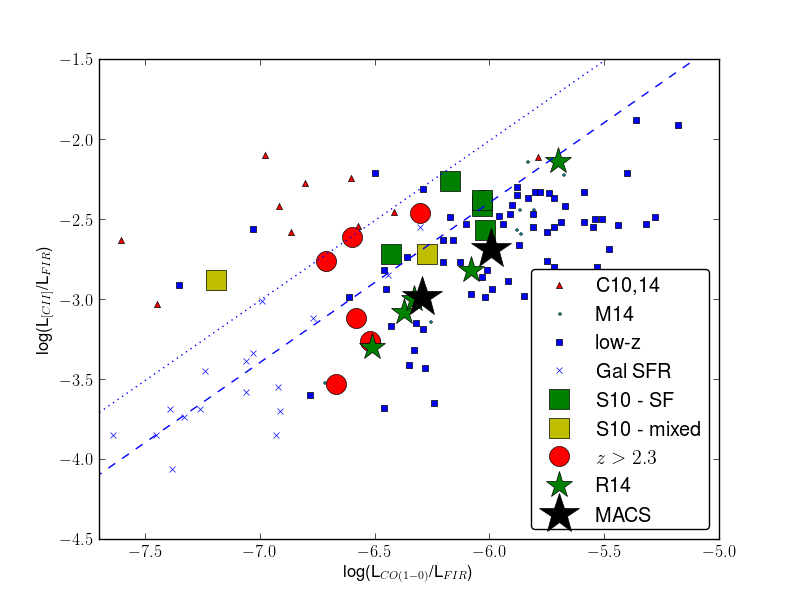}
\caption{\lcii/\lfir\ versus \lco/\lfir\ with \macs\ (large, black stars; cf.\ Fig.\ \ref{fig_cii_lir}) and other samples.
Small symbols show low-$z$ SF galaxies and AGN, large symbols $z>2$ galaxies from the following 
papers:
C10,14: \cite{Cormier2010The-effects-of-}, \cite{Cormier2014The-molecular-g},
M14: \cite{Magdis2014A-Far-Infrared-},
S10: \cite{Stacey2010A-158-mum-C-II-},
R14: \cite{Rawle2014C-II-and-12CO1-}, 
and various $z>2$ galaxies from the literature.
Crosses show Galactic SF regions with low IR luminosities, below the range plotted in Fig.\ \ref{fig_cii_lir}.
The dashed (dotted) line shows a ratio of $\lcii/\lco=4.1 \times 10^3$ ($10^4$).
}
\label{fig_cii_co}
\end{figure}

\subsection{Normal \cii/CO emission}

The \Cii\ and CO line strengths, both normalized to \lfir, are shown in Fig.\ \ref{fig_cii_co}
and compared to available data from nearby and distant galaxies and AGN.
Again \macs\ is found in a region of ``normal" \cii/CO ratios, close to $\lcii/\lco = 4.1 \times 10^3$ shown by the dashed line,
where most nearby starbursts, AGN, and $z \sim 1-2$ star-forming galaxies are found. 
In contrast, nearby quiescent star-forming galaxies show a lower \cii/CO ratio \citep{Stacey2010A-158-mum-C-II-}. 
At the other end, some objects, mostly dwarf galaxies, show significantly higher \cii/CO ratios, exceeding significantly
$\lcii/\lco=10^4$.

Both observations and PDR modeling suggest that high 
\cii/CO ratios are related to low metallicity \cite[cf.][]{Stacey1991The-158-micron-,Cormier2010The-effects-of-,
De-Breuck2011Enhanced-CII-em}. High \cii/CO ratios are explained by spherical PDR models, where the CO emission
region is reduced at low metallicity \citep[see e.g.][]{Bolatto1999A-Semianalytica,Rollig2006CII-158-mum-emi}.
For \macs, \cite{Richard2011The-emission-li} derive a 
metallicity of $12+\log({\rm O/H}) = 8.80^{+0.13}_{-0.12}$ from the \nii/\ha\ ratio.
However, since the region may contain an AGN 
(Zamosjki et al., in preparation),
 \nii/\ha\ may be boosted by the AGN, and the metallicity hence overestimated. 
 Indeed, a low/sub-solar metallicity would be expected from the fairly low
mass of this galaxy. Using, e.g., the fundamental mass-SFR-metallicity relation of \cite{Mannucci2010A-fundamental-r}
one expects $12+\log({\rm O/H}) \approx 8.4$, lower than the above estimate.
However, in comparison to the nearby dwarf galaxies from \cite{Cormier2010The-effects-of-,Cormier2014The-molecular-g},
with metallicities $12+\log({\rm O/H}) = 7.89-8.38$, \macs\ has a higher metallicity, consistent
with a lower \cii/CO ratio.

Compared to most low redshift galaxies (except for low metallicity dwarf galaxies),
\macs\ shows a comparable \lco/\lfir\ ratio (Fig.\ \ref{fig_cii_co}), although a large dispersion is found at all redshifts 
\citep[cf.][]{2010MNRAS.407.2091G,Combes2013Gas-fraction-an,Dessauges-Zavadsky2014Molecular-gas-c}. 
At the corresponding \lir/\lco\ (or equivalently \lir/$M_{\rm H_2}$) ratio, the observed \lcii/\lfir\ ratio
is slightly below, although consistent with the observed trend of local galaxies \citep[][]{Gracia-Carpio2011Far-infrared-Li,Magdis2014A-Far-Infrared-}.

In simple 1D PDR models the main physical parameters are the incident far-UV (FUV) radiation field, commonly 
measured by the Habing flux $G_0$, and the gas density \citep[e.g.][]{Le-Petit2002D/HD-transition}. 
From such models variations of \lco/\lfir\
are mostly explained by varying $G_0$, with an increased FUV flux causing a decrease of  \lco/\lfir\
\citep[cf.][]{Stacey1991The-158-micron-,Stacey2010A-158-mum-C-II-}.
In their $z \sim 1-2$ galaxy sample, \cite{Stacey2010A-158-mum-C-II-} find that galaxies containing an AGN 
have on average a higher FUV flux,
 i.e.\ a lower \lco/\lfir\ ratio. Indeed, \macs\ shows a comparable \lco/\lfir\ ratio as their $z \sim 1-2$ ``mixed" sample,
which could indicate that the ratio is also affected by a presumed AGN contribution in this galaxy.
However, the comparison sample of \cite{Stacey2010A-158-mum-C-II-} is very small, and other galaxies
at high redshift (indicated as $z>2.3$ in Fig.\ \ref{fig_cii_co})  also display comparable properties in the 
\lcii/\lfir\ and \lco/\lfir\ ratios.


\subsection{Discussion}

To place our target into a more general context of IR-detected galaxies, \macs\ is
-- with $\lir=(1.1-1.3) \times 10^{11}$ \lsun\ -- a faint LIRG, with an infrared luminosity $\lir \sim 0.3-0.4 L^\star$, below 
the characteristic value of $L^\star$ at $z=2$ using the luminosity function of \cite{Gruppioni2013The-Herschel-PE},
or at $\sim 0.07 L^{\rm knee}$ using the measurements of \cite{Magnelli2013The-deepest-Her}.
Compared to the stellar mass function of star-forming galaxies at $z \sim 2$ our galaxy has a mass of $\sim 0.031$ $M^\star$ 
\citep[cf.][]{Ilbert2013Mass-assembly-i}.

Compared to other galaxies currently detected in \cii\ at high redshift for which information is
relatively sparse, our rich dataset available for \macs\ allow us to determine quantities
concerning its dust content (mass, temperature, UV attenuation), gas content (CO mass, gas depletion timescale,
gas fraction), stellar content (SFR, mass, approximate metallicity), and
kinematics \citep[see][]{Jones2010Resolved-spectr,Richard2011The-emission-li,Sklias2014Star-formation-,Dessauges-Zavadsky2014Molecular-gas-c}.
From our present knowledge at $z \sim 2$, \macs\ appears as fairly normal for its stellar/star formation properties:
it is, e.g., close to or within the main sequence of \cite{Daddi07}, although this not well determined
in the mass range of \macs.
In terms of ISM/PDR properties, however, only few galaxies are measured at these redshifts,
preventing us from determining what their ``normal"/typical properties are.
In any case, despite a higher sSFR and hotter dust compared to nearby galaxies and possibly the presence 
of an AGN, \macs\ shows relative \Cii/IR and CO/IR properties which are very similar to those of nearby
(and few other $z \sim 1-2$) star-forming galaxies. This shows that the ISM properties depend in a 
more complex manner on several physical parameters.

\section{Conclusion}
\label{s_conclude}

Using ALMA in cycle 0, we have detected \Cii\ emission from the $z=2.013$ strongly lensed, multiply-imaged arc \macs\
previously studied in depth thanks to HST, Spitzer, Herschel, and IRAM observations.
The spatially integrated \cii\ luminosity corresponds to $\lcii = 1.2 \times 10^8$ \lsun, after correction for lensing.
The IR luminosity of this galaxy is $\sim 10$ times fainter than any previous source detected both in \cii\ and in 
the IR continuum at high redshift.
The observed ratio of \cii-to-IR emission, $\lcii/\lfir \approx(1.2-2.4) \times 10^{-3}$, is found to be similar to that
in nearby galaxies. The same also holds for the observed \cii/CO ratio, which is comparable to that of star-forming 
galaxies and AGN at low redshift, and also in agreement with available measurements in more IR-luminous systems
at high redshift.
Although \macs\ shows a high dust temperature \citep[$T_d \ga 50$ K,][]{Sklias2014Star-formation-},
the \lcii/\lfir\ ratio is not lower than in nearby galaxies with comparable dust temperatures 
\citep[cf.][]{Malhotra2001Far-Infrared-Sp,Magdis2014A-Far-Infrared-}.

Our previous CO observations with IRAM and the present \Cii\ detection with ALMA provide a first
hint on PDR/ISM properties of a relatively low mass ($\mstar \sim 2.5 \times 10^9$ \msun) star-forming 
galaxy at $z \sim 2$. 
Observations of larger samples of ``normal" star-forming galaxies at high redshift should soon become
available, providing us then with a better understanding of their ISM and star formation properties.
Further ALMA observations of \macs\ have been approved to study \cii, CO,  and dust emission on small
spatial scales down to $\sim$ 200 pc in the source plane.

\begin{acknowledgements}
This work was supported by the Swiss National Science Foundation.
This paper makes use of the following ALMA data: 2011.0.00130.S (PI Ellis).
ALMA is a partnership of ESO (representing
its member states), NSF (USA) and NINS (Japan), together with NRC
(Canada) and NSC and ASIAA (Taiwan), in cooperation with the Republic of
Chile. The Joint ALMA Observatory is operated by ESO, AUI/NRAO and NAOJ.
We gratefully thank the ALMA staff of NAASC for their assistance in preparing 
the observations, which form the basis of this Letter.
\end{acknowledgements}


\bibliographystyle{aa}
\bibliography{merge_misc_highz_literature}

\end{document}